# MBE growth of self-assisted InAs nanowires on graphene

**Jung-Hyun Kang[1], Yuval Ronen[1], Yonatan Cohen[1], Domenica Convertino[2], Antonio Rossi[2], Camilla Coletti[2], Stefan Heun[3], Lucia Sorba[3], Perla Kacman[4] and Hadas Shtrikman[1]**

[1]Department of Condensed Matter Physics, Braun Center for Submicron Research, Weizmann Institute of Science, Rehovot 76100, Israel
[2]Center for Nanotechnology Innovation at NEST, Istituto Italiano di Tecnologia, Piazza San Silvestro 12, 56127 Pisa, Italy
[3]NEST, Istituto Nanoscienze-CNR and Scuola Normale Superiore, Piazza San Silvestro 12, 56127 Pisa, Italy
[4]Institute of Physics Polish Academy of Science, Al. Lotnikow 32/46, 02-668 Warsaw, Poland

E-mail: hadas.shtrikman@weizmann.ac.il

**Abstract.** Self-assisted growth of InAs nanowires on graphene by molecular beam epitaxy is reported. Nanowires with diameter of ~50 nm and aspect ratio of up to 100 were achieved. The morphological and structural properties of the nanowires were carefully studied by changing the substrate from bilayer graphene through buffer layer to quasi-free-standing monolayer graphene. The positional relation of the InAs NWs with the graphene substrate was determined. A 30° orientation configuration of some of the InAs NWs is shown to be related to the surface corrugation of the graphene substrate. InAs NW-based devices for transport measurements were fabricated, and the conductance measurements showed a semi-ballistic behavior. In Josephson junction measurements in the non-linear regime, Multiple Andreev Reflections were observed, and an inelastic scattering length of about 900 nm was derived.

## 1. Introduction

InAs nanowires (NWs) are key role players in current mesoscopic physics experiments including the search for Majorana fermions [1-3]. This is due to a Fermi level pinning in the conduction band, a large Landé g-factor, a strong spin-orbit coupling in these NWs, and also due to the easy formation of non-alloyed Ohmic contacts. InAs NWs of high crystalline quality and with high aspect ratio are usually grown by the vapor-liquid-solid (VLS) method assisted by a gold droplet either by molecular beam epitaxy (MBE) [4], metal organic chemical vapor deposition (MOCVD) [5, 6], or chemical beam epitaxy (CBE) [7, 8]. The substrate is typically <111>B InAs on which the NWs grow perpendicular to the growth surface. The diameter of the NWs is determined by the diameter of the gold droplet, and their length, for a particular diameter, by the growth time and the arsenic-to-indium ratio. Yet, gold-guided growth imposes a concern over possible incorporation of gold atoms that might act as stutterers [9].

Self-assisted growth of InAs NWs avoids the use of gold or any other foreign metal catalyzers suspected to incorporate during growth. This has attracted quite a bit of attention thanks to the simplicity of the process normally utilizing the most handy/cheap silicon/SiO$_2$ substrate [10-14]. Si/SiO$_2$ substrates are successfully used for growth of III-V NWs, in particular GaAs NWs [15-17]. Self-assisted InAs NWs typically grow with quite prominent [011] facets [18]. The side facets affect the chemical and electronic properties of the NWs surface and are thus important for in-situ metallization [19] and in-situ shell growth [20]. Unfortunately, self-assisted growth on Si/SiO$_2$ normally results in InAs NWs having mixed zinc blende (ZB) / wurtzite (WZ) structure with prominent rotational defects along the growth direction and a typically low aspect ratio [18, 21]. The growth is frequently ascribed to a vapor-solid (VS) mechanism, which is probably responsible for the large number of stacking defects observed in these NWs [18, 21, 22].

The structural compatibility of the honeycomb crystal lattice with the ZB and WZ crystal structures of many III-V semiconductors turns graphene into an attractive platform for the growth of NWs of these materials. Moreover, graphene is a low cost substrate that has high mechanical strength and flexibility, excellent electrical properties and optical transparency. The direct growth of semiconductor NWs on graphene has already been demonstrated, mainly using metal-organic chemical vapor deposition [23-32]. In particular, self-assisted van der Waals growth of InAs NWs on graphene was shown to be an alternative to that on Si/SiO$_2$ [29-32]. A VLS/VS growth mechanism guided by a small persistent indium droplet, at least during the initial stage of the growth, assures the vertical growth, yet often having highly disordered crystal structure and significant side growth [28, 30].

It is well known that molecular beam epitaxy (MBE), which can provide accurate control over the growth parameters, is the best method to produce high-quality NWs. In the last few years Munshi et al. showed that MBE can be used to obtain GaAs NWs on graphene [25, 28]. Zhuang et al. demonstrated In droplet-assisted MBE growth of InAs nanorods on mechanically exfoliated graphite flakes [31], and quite recently Tchoe et al. obtained core/shell InAs/InGaAs coaxial nanorod heterostructures on graphene layers using MBE [32].

The development of a catalyst-free MBE growth method to produce ultrapure InAs NWs with controllable morphological and structural properties is of great importance. In particular, wires with diameters of about 50 nm and high aspect ratio are needed to assure ballistic transport conditions. Also a lower density of NWs, which would allow processing of a device without harvesting the wires, is essential for utilizing the planar conductivity of graphene in semiconductor-graphene hybrid structures.

Here, we report the results of an optimization of self-assisted InAs NW growth on graphene by MBE. The morphological and structural properties of the NWs and their positional relation with the substrate were studied while changing the substrate from bilayer graphene through buffer layer to quasi-free-standing monolayer graphene (QFMLG). In order to characterize the electrical properties of the obtained InAs NWs, normal and Josephson junction NW-devices were fabricated for transport and superconductivity measurements, respectively.

**2. Experimental Details**

*2.1. Growth of Graphene Layers*

Bilayer graphene, buffer layer, and QFMLG samples were obtained on the silicon-terminated face of insulating silicon carbide (4H-SiC (0001)) [CREE Inc.] in a resistively heated cold-wall reactor [Aixtron HT-BM] [33, 34]. First, the SiC (0001) substrates were hydrogen-etched for 5 minutes at a temperature of about 1300 ℃ and a pressure of 450 mbar, in order to obtain atomically flat surfaces [35]. Buffer layer and bilayer graphene were grown via thermal decomposition in argon atmosphere at a pressure of 780 mbar by heating the substrates at 1350 ℃ and 1380 ℃, respectively. In order to obtain QFMLG, the buffer layer samples were later annealed in hydrogen at 900 ℃ [36]. The quality of the samples and the number of layers were assessed by atomic force microscopy and Raman spectroscopy.

*2.2. Growth and Characterization of InAs NWs on Graphene Layers*

InAs NWs were grown via heteroepitaxy by the self-catalyzed MBE technique in a solid source (RIBER-32) system at the standard growth position. The growth conditions were maintained as close as possible for the different substrates used, with only the nucleation conditions being slightly adjusted, e.g., for buffer layer a smaller amount of indium was deposited than on the bilayer (up to half). The graphene samples were grown ex-situ and transferred in air. After annealing the graphene layers overnight at ~450 °C in a dedicated treatment chamber attached to the growth chamber, the InAs NW growth was initiated by deposition of a small amount of In at a substrate temperature of 100°C (indium beam equivalent pressure (BEP) $1\times10^{-8} - 5\times10^{-8}$ for 10-30 seconds). Then, the manipulator was rotated away from the K-cells while the arsenic flux was built up, and the substrate temperature was slowly ramped up (5°C/min) to the growth temperature in the range 450-550°C. For InAs NW growth, an As/In ratio of ~200 was employed. Indium flux was kept the same during the nucleation and growth. When the desired As flux (BEP $~2\times10^{-5}$ Torr) was reached, the manipulator was rotated in front of the K-cells, and the In shutter reopened. In situ reflection high energy electron diffraction (RHEED) shows that the graphene surface remained unchanged during the substrate heating. Emerging of InAs NWs was observed by RHEED pattern seconds after reopening the In shutter.

The morphology of the NWs was investigated using a field-emission scanning electron microscope (FE-SEM) [Zeiss UltraPlus] with images acquired at 45° and 90° tilt for length and diameter measurements, respectively. Structural characterization of the NWs was performed using a transmission electron microscope (TEM) [JEOL JAM-2100]. Samples for TEM and high resolution (HR-) TEM observation were prepared by mechanical transfer of the NWs onto copper TEM grids coated with carbon film.

*2.3. Fabrication of NW devices and Transport Measurements*

For the transport measurements, NWs were removed from the graphene and placed on a pre-fabricated sample by applying sonication and fabrication techniques [37]. The substrate, Si:P$^+$, covered by SiO$_2$ (150 nm thick), served as a back-gate, allowing control of the carrier density in the NWs and the height of the potential barrier. Then, the NW surface was treated by an ammonium polysulfide solution, $(NH_4)_2S_x$. Normal and superconducting contacts were designed using electron beam lithography and deposited in a UHV e-beam system with a composition of Ti-5\Au-120 nm and Ti-5\Al-120 nm, respectively. In the normal contact devices, the NWs were suspended on 30 nm thick Au pillars above the SiO$_2$ substrate.

All transport measurements were carried out in a He3 cryostat at 300 mK using a Lock-In method at a frequency of 571 Hz. The devices were voltage biased up to 10 µV, and the current was amplified ($10^7$) using a home-made current-voltage amplifier [38].

## 3. Results and Discussion

*3.1. InAs Nanowires Grown on Bilayer Graphene and on Buffer Layer*

As can be observed in **Figure 1a**, all InAs NWs grown on bilayer graphene are vertically aligned. Some parasitic islands are also present on the graphene surface. The average aspect ratio of such NWs is, however, rather small, i.e., ~25 (**Figure 1b**) with InAs NW diameter of 100 nm.

In general, the mobility of adatoms diffusing on graphene has been shown to depend strongly on the quality and thickness of the graphene [39, 40]. In particular, the first carbon-rich layer forming on SiC (0001), known as interface or buffer layer, presents a highly corrugated surface that leads to a shorter diffusion length of the adatoms and restrains their mobility. This has been shown to lead to metallic clusters of smaller diameter [39]. Thus, in an attempt to improve the aspect-ratio of the InAs NWs, a homogenous $(6\sqrt{3} \times 6\sqrt{3})$ R30°-reconstructed buffer layer surface was used as a substrate. As shown in **Figures 1c-d,** indeed, by switching to the buffer layer substrate and adjusting the nucleation conditions, InAs NWs with significantly smaller diameter of ~35-50 nm and an improved aspect-ratio (75-100) were obtained. This could be related to an enhanced 2D growth on the buffer layer due to its higher corrugation. Moreover, a more uniform distribution of diameters has been attained. Finally, possible evidence for the growth being guided by indium droplets, which are transformed into InAs at the end of the growth, can be deduced from the rounded NW tips (**Figure 1e**). Potts et al, Zhuang et al, and Hertenberger et al. observed a similar shape at the tip of their MBE-grown InAs NWs which they considered to be clear evidence that in their case growth proceeds without droplet [22, 31, 41]. We would like to comment that we have a vast experience growing self-assisted GaAs nanowires on Si under a wide range of growth conditions. Yet contrary to other groups [42, 43] we were never able to show the existence of a droplet post growth. We rather found a round shape, very similar to the one occasionally seen for self-assisted InAs nanowires, but quite different from the flat shape related to the VS mechanism [21]. Since the self-assisted growth of InAs nanowires is carried out at a particularly high As/In ratio. The chance of maintaining a liquid droplet as a proof for the growth being VLS is negligible. Thus, at best one can expect to see a round shape which is the remanence of the droplet after the In shutter has been closed (even when closing both shutters at the same time).

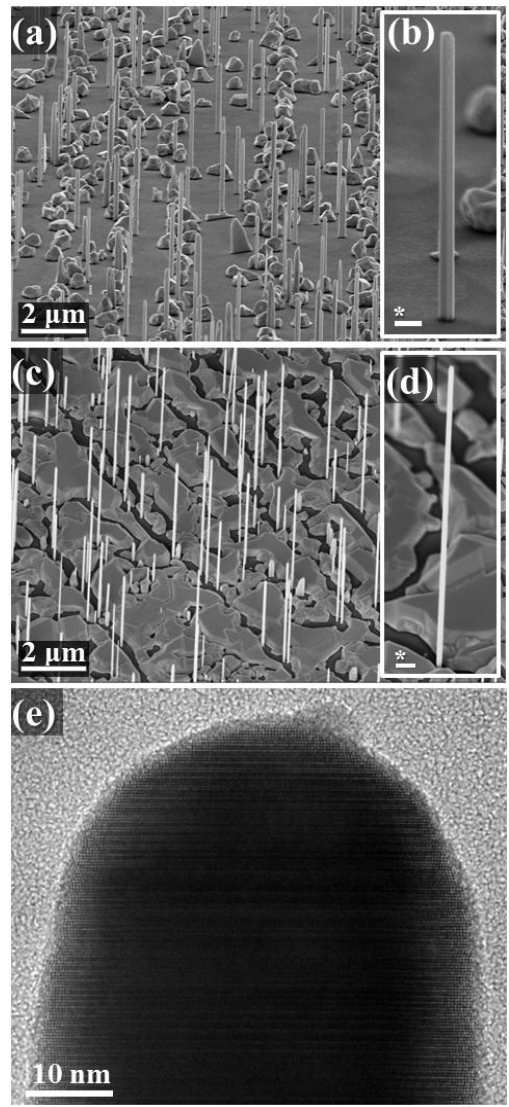

**Figure 1.** **(a)** Bird-eye view and **(b)** enlarged SEM micrographs of a single InAs NW grown on bilayer graphene (average aspect ratio ~25); **(c)** bird-eye view and **(d)** enlarged SEM images of a single InAs NW grown on buffer layer (average aspect ratio ~100), showing enhanced 2D growth on the buffer layer; **(e)** HR-TEM image of the tip of a NW grown on the buffer layer; The *scale bars are 100 nm.

The crystal structure of InAs NWs grown on bilayer graphene (**Figure 2a**) and on the buffer layer (**Figure 2b**) has been investigated by HR-TEM measurements. As can be seen in **Figure 2**, the HR-TEM images reveal a mixed WZ and ZB crystal structure of the InAs NWs on both substrates. The numerous WZ/ZB stacking faults are most probably a result of the typically low growth temperature of InAs NWs (~550°C). This causes a low indium surface tension and low vapor pressure at the droplet surface as compared to the growth conditions used for growth of twin plane free GaAs NWs, i.e., high temperature, ~650 °C and low Ga flux [15].

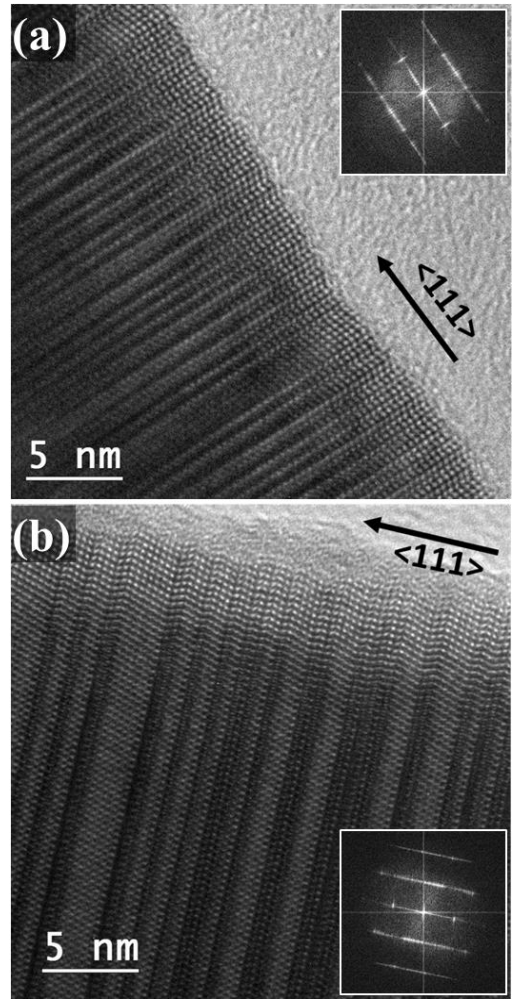

**Figure 2.** HR-TEM micrographs and FFT images (inset) of InAs NWs grown on **(a)** bilayer graphene and **(b)** buffer layer. The TEM images and their respective FFTs both show a mixed ZB/WZ structure.

*3.2. The Orientation of InAs Nanowires Relative to the Substrate*

It was discussed by Munshi et al. [in Ref. 25] that in the case of some semiconductors the lattice mismatch with graphene is very small for only one suggested atomic configuration. As an example the authors discuss InAs, for which only the orientation relation, as sketched in Figure 1c in their paper (which they called 0° configuration), is expected. This has been indeed experimentally observed by Hong et al. [44]. By analyzing plan-view SEM micrographs we have also observed that InAs NWs grown on bilayer graphene were mostly aligned with the graphene lattice. However, over a large area we did see occasional rotations of a single NW by 30° around the growth axis.

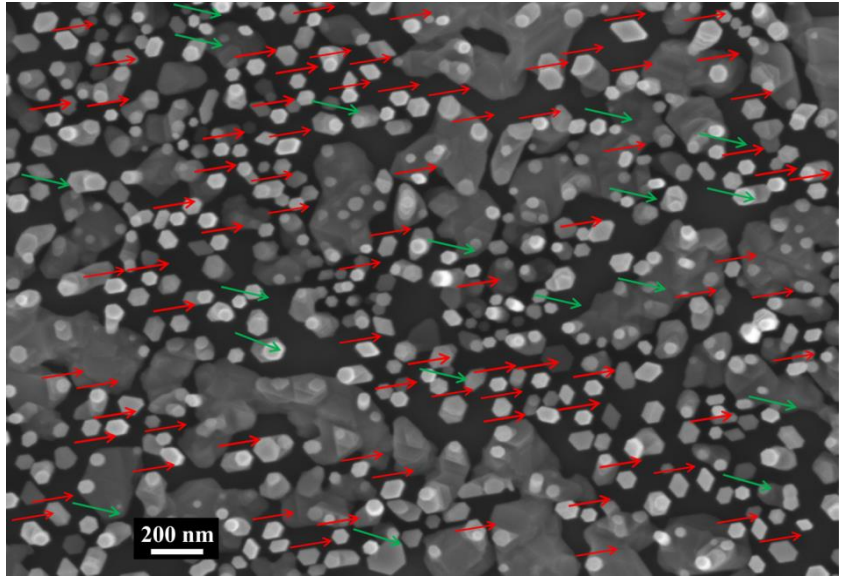

**Figure 3.** Plan-view SEM micrograph of InAs NWs grown on a buffer layer; red and green arrows indicate NWs rotated by 30° around the growth axis.

Interestingly, the top view SEM images show that InAs NWs grown on buffer layer exhibit significantly more rotations by 30°, as can be seen in **Figure 3**. The rotated NWs are homogeneously distributed all over the surface. One direction is indeed quite dominant but the other is prominently present - here we observe well over 10% of InAs NWs rotated by 30° with respect to the others. The 30° rotation is very puzzling, because the hexagonal shape of the NWs grows with an epitaxial relationship with the graphene lattice, which in turn has a 60° rotational symmetry. The reasons behind these 30° rotations are thus not obvious. One might wonder whether imperfections in the graphene substrates are responsible for the observed unexpected rotation of some wires. Significant polycrystallinity of the graphene film can be, however, excluded, because on the Si-face of SiC, graphene grows with a well-defined epitaxial relationship to the (single crystal) SiC substrate, as demonstrated by a bright diffraction pattern in low-energy electron diffraction (LEED) [45-50]. In particular, no graphene-related diffraction spots at 30° angles are observed in the LEED patterns. Furthermore, our scanning tunneling microscope (STM) investigations on buffer layer and monolayer graphene on the Si-face of SiC never showed any evidence for the presence of 30° domains [33, 51].

Next, one can speculate that a surface partially covered with graphene and partially displaying uncovered SiC might explain the observed 30° rotation of some NWs. This reasoning is based on the well-known fact that graphene grows with a unit cell which is 30° rotated with respect to the unit cell of the underlying SiC substrate (**Figure 4**). However, the presence of a significant fraction of uncovered SiC substrate can be excluded here, because micro-Raman spectroscopy on bilayer graphene samples unequivocally yields graphene-related spectra *everywhere* on the sample. Moreover, even the existence of uncovered SiC in the substrate would not explain the rotation, because in a control experiment we have verified that InAs NWs do not grow on SiC (0001) substrates. Finally, also monolayer (or trilayer) inclusions in bilayer graphene will not result in the appearance of 30° domains, because graphene layers grown on the Si-face of SiC maintain the epitaxial relationship with the SiC substrate, i.e. the graphene hexagons in the different layers are all azimuthally oriented.

Another possible explanation for the observation of 30° rotated NWs could be that the real lattice constants (of graphene and/or InAs) deviate from the tabulated room temperature values due to strain or different thermal expansion coefficients. In order to obtain NWs rotated by 30° (**Figure 1b** in Ref. [25]), the lattice constant of InAs (~6 Å) would need to move towards the lattice constant of the 30° configuration (~5.25 Å), or vice versa. To check whether such change of lattice constants is possible, we consider the influence of both the thermal expansion coefficients and strain.

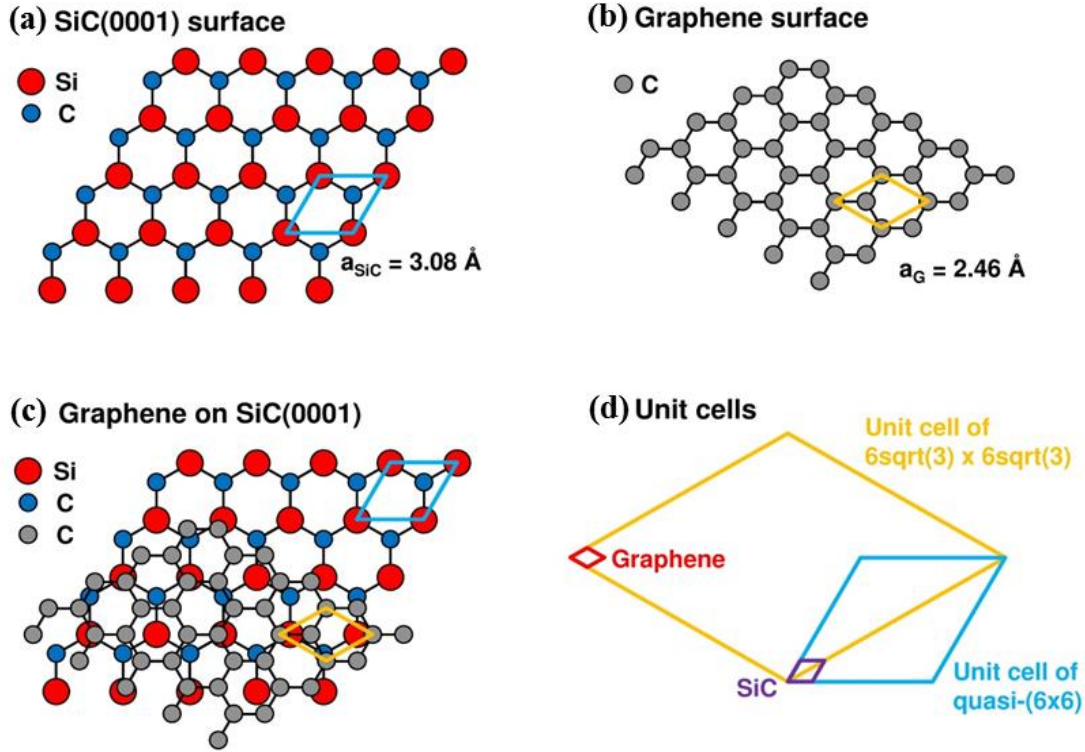

**Figure 4. (a)** Sketch of the SiC (0001) lattice with unit cell. **(b)** Sketch of the graphene lattice with unit cell. **(c)** Sketch of graphene grown on top of SiC (0001), showing that the two unit cells are rotated by 30° with respect to each other. **(d)** Dimensions and orientation of the unit cells and reconstructions discussed in the text. Red: Graphene unit cell, $a_G$ = 2.46 Å; violet: SiC (0001) unit cell, $a_{SiC}$ = 3.08 Å; light blue: unit cell of quasi-(6x6), size: $6 \times a_{SiC}$ = 18.48 Å; yellow: unit cell of $(6\sqrt{3} \times 6\sqrt{3})$ R30°, size: $6\sqrt{3} \times a_{SiC}$ ~ $13 \times a_G$ ~ 32 Å.

**Table 1**. Lattice constants of the buffer layer and of bilayer graphene for configurations 30° and 0° (defined in **Figure 1** of Ref. [25]) compared to the lattice constant of InAs, both for room temperature (RT) and for 400 °C.

|  |  | Lattice Constant, $a$ (Å) | | |
|---|---|---|---|---|
|  |  | Configuration 30° | Configuration 0° | InAs |
| RT | Buffer Layer | 5.224 | 6.033 | 6.0583 |
|  | Bilayer Graphene | 5.224 | 6.033 |  |
| 400 °C | Buffer Layer | 5.230 | 6.040 | 6.0693 |
|  | Bilayer Graphene | 5.207 | 6.014 |  |

If we assume that due to the strong covalent bonding to the SiC substrate the buffer-layer has the same thermal expansion coefficient as SiC ($\alpha_{SiC}$ = 2.77 × 10$^{-6}$ K$^{-1}$), while bilayer graphene has the thermal expansion coefficient of graphene, which is negative, $\alpha_G$ = -8.0 × 10$^{-6}$ K$^{-1}$), and for InAs $\alpha_{InAs} = 4.52 \times 10^{-6}$ K$^{-1}$, we obtain the values summarized in **Table 1**. Evidently, this effect leads only to small variations in lattice constant, and for both the buffer layer and bilayer graphene, the InAs is locked to the 0° configuration.

Similarly, the change of the lattice constants caused by strain cannot account for the observed 30° rotation of the NWs. For the buffer layer, a compressive strain of about 0.8 % at room temperature has been reported [52], caused by the lattice mismatch between graphene and SiC. Strain in the bilayer is expected to be lower. A ±1% strain would lead to the following lattice constant ranges: [5.172 Å, 5.276 Å] for the 30° configuration, [5.973 Å, 6.093 Å] for the 0° configuration, and [5.998 Å, 6.119 Å] for InAs. Again, the lattice constant of InAs is locked to the 0° configuration. Thus, we can conclude that variations in lattice constant due to strain or/and different thermal expansion cannot account for the observed 30° rotation of NWs.

### 3.2.1. Perfect Alignment of InAs Nanowires Grown on Quasi-Free Standing Monolayer Graphene – the corrugated structure of the buffer layer explains the rotation of the NWs

We finally relate the observation of a 30° rotation of some NWs to the corrugated structure of the buffer layer. **Figure 4** illustrates this point. The SiC (0001) surface has a hexagonal structure with a lattice constant of 3.08 Å (**Figure 4a**). Graphene also has a hexagonal structure with a lattice constant of 2.46 Å (**Figure 4b**). Graphene on SiC forms a commensurate superstructure, with the graphene unit cell rotated by 30° with respect to the SiC unit cell, and a periodicity of 13 × 13 graphene unit cells or $(6\sqrt{3} \times 6\sqrt{3})$ R30° with respect to the SiC lattice – in fact, 13 × 2.46 Å ~ $6\sqrt{3} \times 3.08$ Å ~ 32 Å (see **Figure 4d** and Ref. [33]). This superstructure is associated with a periodic corrugation of the surface (Moiré pattern), caused by partial bonding of C atoms to the underlying SiC substrate: about 30 % of the carbon atoms constituting the buffer layer are covalently bonded to the Si atoms of the SiC (0001) substrate [53]. The corrugation is strongest for the buffer layer, but observed also on monolayer and bilayer graphene due to the interaction with the underlying buffer layer (the corrugation diminishes with increasing number of graphene layers). STM investigations show that the height variation of the surface follows approximately a quasi-(6 × 6) reconstruction, which is rotated by 30° with respect to the graphene lattice and aligned with the SiC lattice (see **Figure 4d** and Ref. [33]). We suggest that the InAs growth in some cases follows the quasi-(6 × 6) unit cell, in a kind of indirect epitaxy [54], i.e., the NWs grow aligned with the surface corrugation. Similar ordering effects have already been observed on corrugated graphene [55]. The much stronger surface corrugation on the buffer layer explains also why we see more rotated NWs on the buffer layer as compared to bilayer graphene. The quasi-(6×6) unit cell has a periodicity of 18.48 Å (6 × 3.08 Å, which is the SiC lattice constant). If we compare this with the InAs lattice constant in the (111) plane, 6.0583 Å / $\sqrt{2}$ = 4.284 Å, we see that approximately 4-InAs unit cells would fit into the quasi-(6 × 6) template: 4 × 4.284 Å = 17.1 Å. We anticipate that the remaining lattice mismatch of ~7 % can be easily accommodated by the NWs, like for the case of GaAs NWs [25].

In order to test the correctness of this hypothesis we carried out InAs NW growths on QFMLG. QFMLG is obtained by intercalating hydrogen atoms between the buffer layer and the underlying SiC substrate (see **Figure 5a**), thus breaking the existing covalent bonds and saturating them with hydrogen atoms [36]. As a result the graphene layer is structurally decoupled from SiC and literally floating atop the substrate. Atomic resolution STM images of the QFMLG have, in fact, revealed that the periodic corrugation is completely removed, and a perfectly flat surface is realized [33]. The growths of InAs NWs on QFMLG have resulted in fully aligned NWs; see **Figure 5b**, where a negligible number of rotated wires can be seen in a larger area. This result confirms that the growth aligned with the surface corrugation is responsible for the presence of NWs rotated by 30°.

### 3.3. Electrical Properties of InAs NW Devices

With respect to using NWs grown on graphene as building blocks of possible future InAs NWs-based electronic devices, it is important to investigate their transport properties. We first present transport measurements on an InAs NW grown on the buffer layer: device-1 (**d1**), showing a semi-ballistic behavior. Next, a Josephson junction measurement has been performed: device-2 (**d2**), which allowed observing the Multiple Andreev Reflection (MAR) effect in the non-linear regime.

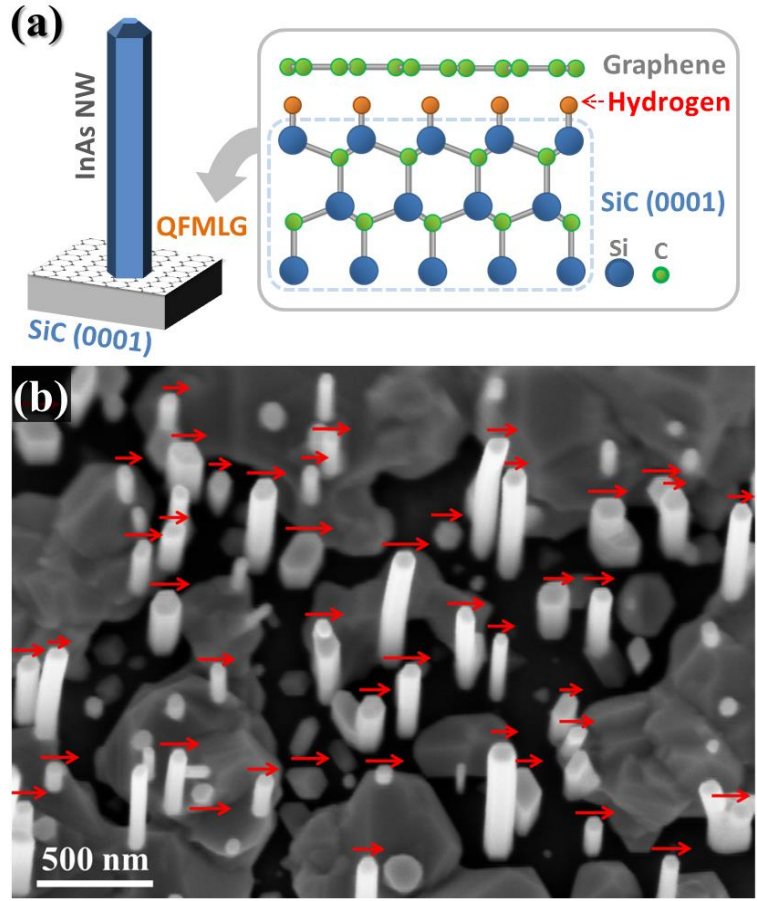

**Figure 5.** (**a**) A schematic illustration and (**b**) a SEM image of InAs NWs grown on quasi-free standing monolayer graphene (QFMLG), showing a 100 % alignment of the InAs hexagons with the graphene lattice. Fuzzy SEM image is a result of the fact that the QFMLG layer is floating and easily charged.

*3.3.1. Transport Measurement of a Single-NW Device with Applied Magnetic Field*

We begin with **d1** which is a suspended device with Ohmic contacts defining a channel length ($L_1$) of 175 nm (**Figure 6a**); NW diameter is 110 nm. In **Figure 6b**, we show the differential conductance of **d1** as a function of back gate voltage, measured at 300 mK and zero magnetic field as well as under a perpendicular magnetic field of 5 Tesla (T) (in the direction described in the inset). Applying magnetic field (red-solid line) reduces Fabry-Pérot conductance oscillations of the signal based on contact resistivity [56] in comparison with the zero magnetic field behavior (grey-dotted line), an effect seen in many samples. Therefore, the magnetic field can be thought as minimizing backscattering effects either in the NW itself or in the contacts. Moreover, when a magnetic field of 5 T is applied (red), clear conductance plateaus are observed at ~1.5 $e^2/h$ and ~2.4 $e^2/h$ (red arrows), and possibly also at 0.7 - 0.75 $e^2/h$.

In the ballistic regime, conductance plateaus should be seen at integer multiples of $2e^2/h$, in the absence of magnetic field. In the presence of a strong magnetic field, due to the Zeeman effect, the energy bands are split, and conductance plateaus are expected to form at integer multiples of $e^2/h$ [57, 58]. However, contact resistance can shift the plateaus from their ideal values. Moreover, since the contact resistance in these devices strongly depends on the back gate voltage, the values of the plateaus deviate from the ideal values, and clear conclusions are difficult to obtain. Furthermore, effects such as electrons back-scattering from stacking faults or twin-defects (**Figure 2**) and potential barriers coming from NW surface contamination can change the conductance values, as well. It is also important to note that many body effects can also play a role and give rise to conductance plateaus [59]. It is therefore difficult to conclude that the channel is indeed ballistic. However, it is important to note the emergence of the conductance plateaus, since similar devices fabricated using InAs NWs grown on the typical InAs substrate usually do not exhibit such plateaus. Finally, we note that in InAs NWs grown on InAs the Landé g-factor of ~18 (about twice as large as InAs bulk values) has been observed [2]. This increase is ascribed mainly to the confinement. Therefore, we expect similar values of the Landé g-factor in InAs NWs studied here. At B = 5 T such Landé g-factor would give rise to a Zeeman energy of ~5 meV and can explain the observed plateau separation.

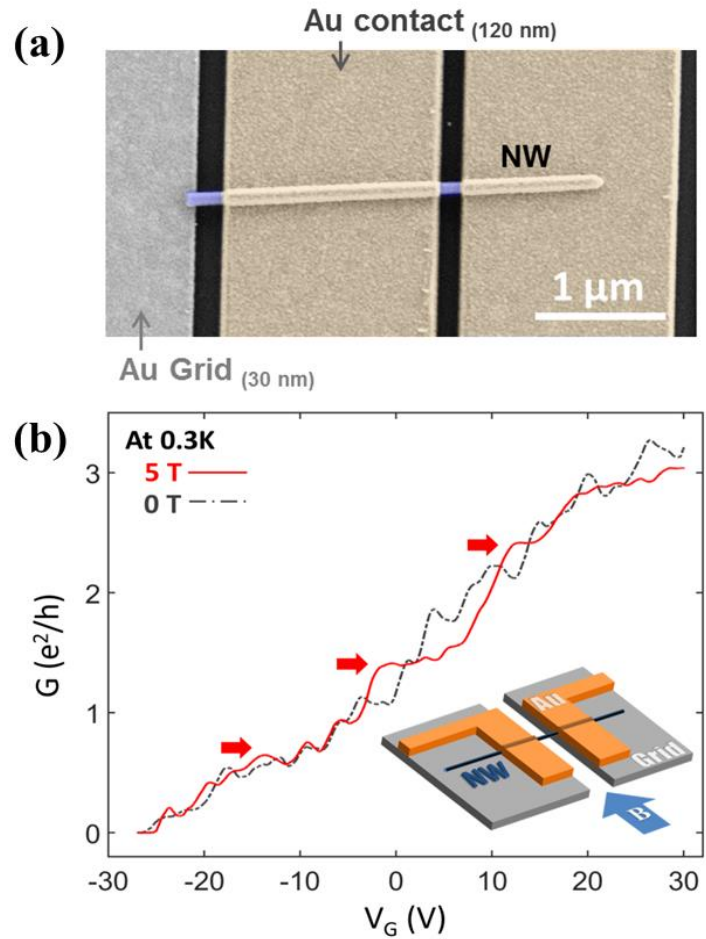

**Figure 6. (a)** SEM image showing device **d1** of a single InAs NW which was grown on the buffer layer. **(b)** Differential conductance G = $\Delta I/\Delta V$ as a function of gate voltage $V_G$ with zero (grey-dotted line) and 5 Tesla (red-solid line) magnetic field parallel to the substrate and perpendicular to the NW, as shown in the schematic image of the device with magnetic field (inset).

*3.3.2. Superconductivity of the Single-NW device and Multiple Andreev Reflection*

We turn to discuss **d2**, where an unsuspended InAs NW grown on QFMLG was contacted by two superconducting contacts to create a Josephson junction with a channel length ($L_2$) of 300 nm and a NW diameter ($D_2$) of 110 nm (shown in **Figure 7a**) [60-62]. In **Figure 7b**, we show the differential conductance as a function of source-drain voltage on the junction ($V_{SD}$) as well as the back-gate voltage ($V_g$). A pronounced subharmonic gap structure is seen, indicating that the so called Multiple Andreev Reflections (MAR) dominate the transport at bias voltages below $2\Delta/e$ [63], where $\Delta = 150$ $\mu$eV. In MAR, quasiparticles can flow from one superconductor to the other by traversing the junction multiple times, undergoing an Andreev reflection each time, until they can escape to the continuum [63]. This gives rise to differential conductance peaks at $eV_{SD} = 2\Delta/n$, where $n$ is an integer.

**Figures 7c** and **7d** show two cuts of **Figure 7b** along constant gate voltage lines. The first cut (**Figure 7c**) shows the conductance vs. $V_{SD}$ for $V_g = 28$V. High sub-gap conductance is seen with clear MAR peaks at $eV_{SD} = 2\Delta/n$ for $n = 1, 2$, and 3 and a strong supercurrent peak at zero bias stating a coherent transport along the NW. The second cut (**Figure 7d**) shows the conductance vs. $V_{SD}$ at $V_g = -0.2$V. At this gate voltage the single particle transmission probability, $t$, is very low. From the normal conductance, measured at $eV_{SD} > 2\Delta$, we estimate the transmission probability at $t = G/G_0 = (0.2\ e^2/h)/(2\ e^2/h) = 0.1$ (assuming a single channel occupation in the NW) [64]. This strongly suppresses the sub-gap current since any $n$-th order MAR process is suppressed by $t^n$ (the quasiparticle traverses the junction $n$ times) [64]. In addition, one can extract a lower bound for the inelastic scattering length according to the highest MAR order [65]. In the high gate voltage configuration (**Figure 7c**), the $3^{rd}$ MAR peak is clearly observable (at $eV_{SD} = 2\Delta/3$). Thus, this gives lower bound for the inelastic scattering length of $3L_2$ ($3 \times 300$ nm) $= 900$ nm.

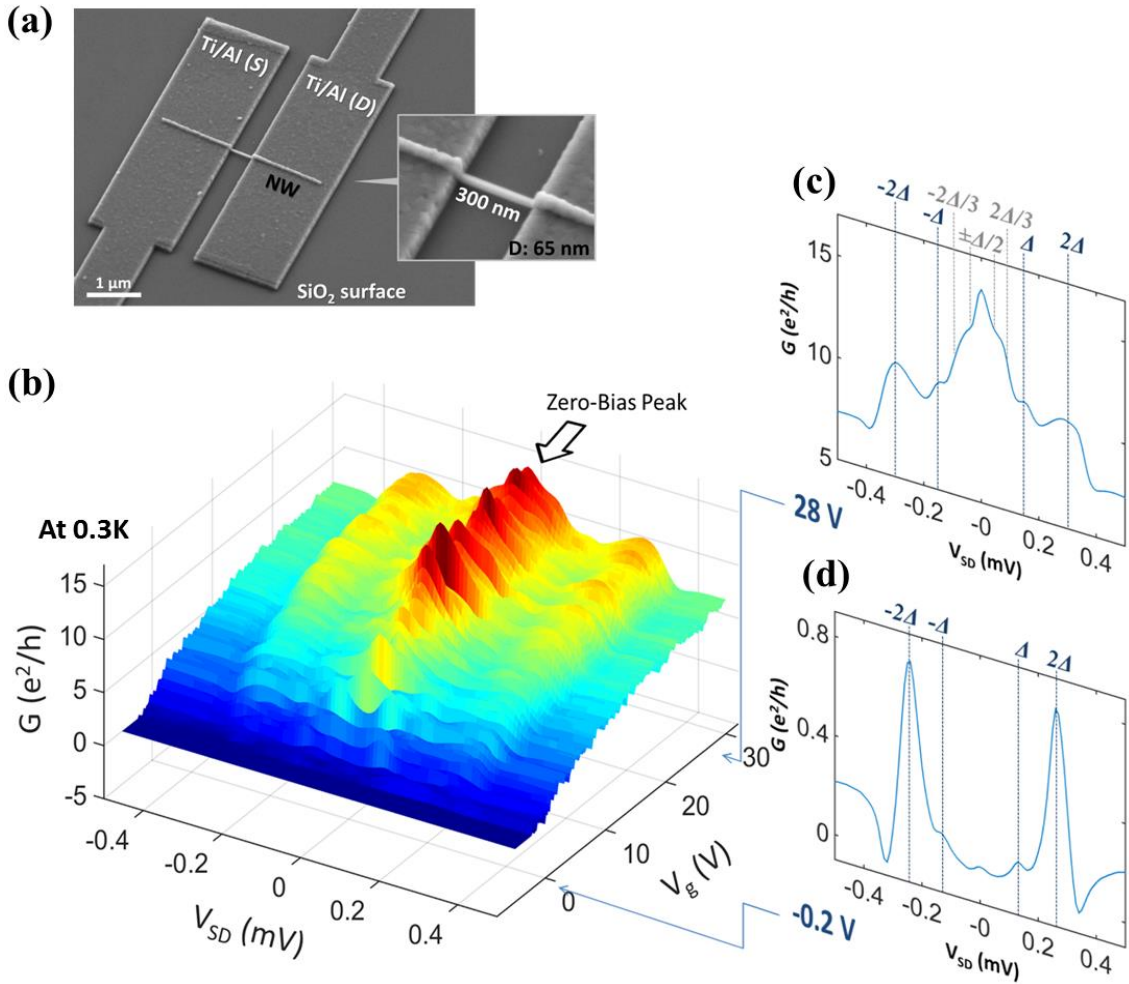

**Figure 7.** (a) Josephson junction device d2 of a single InAs NW grown on QFMLG layer and (b) differential conductance (G) as a function of source-drain voltage ($V_{SD}$) and back gate voltage ($V_g$); (c-d) two cuts of (b) for constant back gate voltages $V_g = 28$ V and -0.2 V, respectively.

## 4. Conclusions

In this study, we have shown that the morphological properties of self-catalyzed InAs NWs grown by MBE can be considerably improved by a proper choice of the substrate. It has been shown that the buffer layer (graphene/SiC interface) is the best substrate for such growth. Indeed, on reconstructed buffer layer surfaces, InAs NWs with optimized aspect ratio (75-100), small diameters (30-50 nm) and a uniform distribution were obtained.

Surprisingly enough, for over 10% of these NWs we observed a 30° rotation along the growth axis. We suggest that this 30° rotation is due to the enhanced periodic surface corrugation of the buffer layer. We demonstrated that eliminating the corrugation (by separating the graphene layer from the SiC substrate by intercalated hydrogen atoms) yields perfectly aligned InAs NWs. Low temperature transport measurements demonstrate good electric properties of these NWs; namely, conductance steps upon applying magnetic field and superconductivity with symmetric Multiple Andreev Reflection peaks could be observed. The results presented pave the way for using InAs NWs on graphene for future electronic device implementations.

**Acknowledgments**

The Israeli Science Foundation and Israeli Ministry of Science and Technology are acknowledged for partial financial support (Grant No. 532/12 and Grant No. 3-6799). We also thank the European Union Seventh Framework Programme under grant agreement no. 604391 Graphene Flagship; the CNR in the framework of the agreements on scientic collaborations between CNR and JSPS (Japan), CNRS (France), NRF (Korea), and RFBR (Russia); and the Italian Ministry of Foreign Affairs, Direzione Generale per la Promozione del Sistema Paese and the Polish National Science Center grant No. 2013/11/B/ST3/03934. The authors are grateful to Moty Heiblum for making this research possible. We are grateful to Michael Fourmansky for his unstinting technical assistance. Discussions with Stefano Guiducci are gratefully acknowledged. Perla Kacman acknowledges working in WIS supported by the Erna and Jacob Michael Visiting Professor fellowship.

**References**

[1] Lutchyn R M, Sau J D and Das Sarma S. 2010 *Phys. Rev. Lett.* **105** 077001
[2] Das A, Ronen Y, Most Y, Oreg Y, Heiblum M and Shtrikman H. 2012 *Nat. Phys.* **8** 887
[3] Mourik V, Zuo K, Frolov S M, Plissard S R, Bakkers E P A M and Kouwenhoven L P 2012 *Science* **336** 1003–1007
[4] Tchernycheva M, Travers L, Patriarche G, Glas F, Harmand J C, Cirlin G E and Dubrovskii V G 2007 *J. Appl. Phys.* **102** 094313
[5] Caroff P, Dick K A, Johansson J, Messing M E, Deppert K and Samuelson L 2009 *Nat. Nanotechnology* **4** 50-55
[6] Dick K A, Deppert K, Mårtensson T, Mandl B, Samuelson L and Seifert W 2005 *Nano Lett.* **5** 761-764
[7] Jensen L E, Björk M T, Jeppesen S, Persson A I, Ohlsson B J and Samuelson L 2004 *Nano Lett.* **4** 1961-1964
[8] Gomes U P, Ercolani D, Zannier V, Beltram F and Sorba L 2015 *Semicond. Sci. Tech.* **30** 010301-014013
[9] Bar-Sadan M, Barthel J, Shtrikman H and Houben L 2012 *Nano Lett.* **12** 2352-2356
[10] Hertenberger S, Rudolph D, Bichler M, Finley J J, Abstreiter G and Koblmüller G 2010 *J. Appl. Phys.* **108** 114316
[11] Mandl B, Stangl J, Mårtensson T, Mikkelsen A, Eriksson J, Karlsson L S, Bauer G, Samuelson L and Seifert W 2006 *Nano Lett.* **6** 1817-1821
[12] Madsen M H, Aagesen M, Krrogstrup P, Sørensen C, and Nygård J 2011 *Nanoscale Research Letters* **6** 516
[13] Koblmüller G, Hertenberger S, Vizbaras K, Bichler M, Bao F, Zhang J P and Abstreiter G 2010 *Nanotechnology* **21** 365602

[14] Gomes U P, Ercolani D, Sibirev N V, Gemmi M, Dubrovskii V G, Beltram F and Sorba L 2015 *Nanotechnology* **26** 415604
[15] Krogstrup P, Pobovitz-Biro R, Johnson E, Madsen M H, Nygård J and Shtrikman H 2010 *Nano Lett.* **10** 4475-4482
[16] Paek J H, Nishiwaki T, Yamaguchi M and Sawaki N 2009 *Phys. Status Solidi C* **6** 1436-1440
[17] Plissard S, Dick K A, Wallart X and Caroff P 2010 *Appl. Phys. Lett.* **96** 121901
[18] Grap Th, Rieger T, Blömers Ch, Schäpers Th, Grützmacher D and Lepsa M I 2013 *Nanotechnology* **24** 335601
[19] Krogstrup P, Ziino N L B, Chang W, Albrecht S M, Madsen M H, Johnson E, Nygård J, Marcus C M and Jespersen T S 2015 *Nat. Mat.* **14** 400-406
[20] Rieger T, Grützmacher D and Lepsa M I 2015 *Nanoscale* **7** 356-364
[21] Dimakis E, Lähnemann J, Jahn U, Breuer S, Hilse M, Geelhaar L and Riechert H 2011 *Cryst. Growth Des.* **11** 4001-4008
[22] Potts H, Friedl M, Amaduzzi F, Tang K, Tütüncüoglu G, Matteini F, Lladó E A, McIntyre P C and Fontcuberta i Morral A 2016 *Nano Lett.* **16** 637-643
[23] Mohseni P K, Behkan A, Wood J D, Zhao X, Yu K J, Wang N C, Rockett A, Rogers J A, Lyding J W, Pop E and Li X 2014 *Adv. Mater.* **26** 3755-3760
[24] Hong Y J, Yang J W, Lee W H, Ruoff R S, Kim K S and Fukui T 2013 *Adv. Mater.* **25** 6847-6853
[25] Munshi A M, Dheeraj D L, Fauske V T, Kim D C, van Helvoort A T J, Fimland B O and Weman H 2012 *Nano Lett.* **12** 4570-4576
[26] Tateno K, Takagi D, Zhang G, Gotoh H, Hibino H and Sogawa T 2012 *Mater. Res. Soc. Symp. Proc.* **1439** 45-50
[27] Park J B, Kim N J, Kim Y J, Lee S H and Yi G C 2014 *Curren. Appl. Phys.* **14** 1437-1442
[28] Munshi A M and Weman H 2013 *Phys. Status Solidi RRL* **7,** 713-726
[29] Hong Y J and Fukui T 2011 *ACS NANO* **5** 7576-7584
[30] Mohseni P K, Behkan A, Wood J D, English C D, Lyding J W, Pop E and Li X 2013 *Nano Lett.* **13** 1153-1161
[31] Zhuang Q D, Anyebe E A, Sanchez A M, Rajpalke M K, Veal T D, Zhukov A, Robinson B J, Anderson F, Kolosov O and Fal'ko V 2014 *Nanoscale Research Letter* **9** 321
[32] Tchoe Y, Jo J, Kim M and Yi G C 2015 *NPG Asia Mater.* **7** e206
[33] Goler S, Coletti C, Piazza V, Pingue P, Colangelo F, Pellegrini V, Emtsev K V, Forti S, Starke U, Beltram F and Heun S 2013 *Carbon* **51** 249
[34] Bianco F, Perenzoni D, Convertino D, De Bonis S L, Spirito D, Perenzoni M, Coletti C, Vitiello M S and Tredicucci A 2015 *Apl. Phys. Lett.* **107** 131104
[35] Frewin C L, Coletti C, Riedl C, Starke U and Saddow S E 2009 *Mater. Sci. Forum* **589** 615-617
[36] Riedl C, Coletti C, Iwasaki T, Zakharov A A and Starke U 2009 *Phys. Rev. Lett.* **103** 246804
[37] Kang J H, Cohen Y, Ronen Y, Heiblum M, Buczko R, Kacman P, Popovitz-Biro R and Shtrikman H 2013 *Nano Lett.* **13** 5190-5196
[38] Kretinin A V and Chung Y 2012 *Review of Scientific Instruments* **83** 084704
[39] Poon S W, Chen W, Tok E S and Wee A T S 2008 *Appl. Phys. Lett.* **92** 104102
[40] Hardcastle T P, Seabourne C R, Zan R, Brydson R M D, Bangert U, Ramasse Q M, Novoselov K S and Scott A J 2013 *Phys. Rev. B* **87** 195430
[41] Hertenberger S, Rudolph D, Bolte S, Paosangthong W, Spirkoska D, Finley J, Abstreiter G and Koblmueller G 2011 *Appl. Phys. Lett.* **98** 123114
[42] Bauer B, Rudolph A, Soda M, Fontcuberta i Moral A, Zweck J, Schuh D and Reiger E 2010 *Nanotechnology* **21** 435601
[43] Plissard S, Dick K A, Larrieu G, Godey S, Addad A, Wallart X and Caroff P 2010 *Nanotechnology* **21** 385602
[44] Hong Y J, Lee W H, Wu Y, Ruoff R S and Fukui T 2012 *Nano Lett.* **12** 1431
[45] Forti S and Starke U 2014 *J. Phys. D* **47** 094013
[46] Hass J, de Heer W A and Conrad E H 2008 *J. Phys.: Condens. Matter* **20** 323202

[47] Berger C, Song Z, Li T, Li X, Ogbazghi A Y, Feng R, Dai Z, Marchenkov A N, Conrad E H, First P N and de Heer W A 2004 *J. Phys. Chem. B* **108** 19912-19916
[48] Riedl C, Coletti C and Starke U 2010 *J. Phys. D: Appl. Phys.* **43** 374009
[49] Seyller Th, Bostwick A, Emtsev K V, Horn K, Ley L, McChesney J L, Ohta T, Riley J D, Rotenberg E and Speck F 2008 *Phys. Stat. Sol.(b)* **245** 1436-1446
[50] Xia C, Watcharinyanon S, Zakharov A A, Yakimova R, Hultman L, Johansson L I and Virojanadara C 2012 *Phys. Rev. B* **85** 045418
[51] Goler S, Coletti C, Tozzini V, Piazza V, Mashoff T, Beltram F, Pellegrini V and Heun S 2013 *J. Phys. Chem. C* **117** 11506−11513
[52] Ferralis N, Maboudian R and Carraro C 2008 *Phys. Rev. Lett.* **101** 156801

[53] Emtsev K V, Speck F, Seyller Th, Ley L and Riley J D 2008 *Phys. Rev. B* **77** 155303
[54] Cahn R W 1995 *Nature* **375** 363-364
[55] Wang B, Yoon B, König M, Fukamori Y, Esch F, Heiz U and Landman U 2012 *Nano Lett.* **12** 5907
[56] Kretinin A V, Popovitz-Biro R, Mahalu D and Shtrikman H 2010 *Nano Lett.* **10** 3439-3445
[57] Paschen F and Back E 1921 *Physica* **1** 261–273
[58] Weperen I Vsn, Plissard S R, Bakkers E P A M, Frolov S M and Kouwenhoven L P 2013 *Nano Lett.* **13** 387-391
[59] Heedt S, Prost W, Schubert J, Grützmacher D and Schäpers Th 2016 *Nano Lett.* **16** 3116-3123
[60] Buitelaar M R, Belzig W, Nussbaumer T, Babic B, Bruder C and Schoenenberger C 2003 *Phys. Rev. Lett.* **91** 057005
[61] Hoss T, Strunk C, Nussbaumer T, Huber R, Staufer U and Schoenenberger C 2000 *Phys. Rev. B* **62** 4079-4085
[62] Nilsson H A, Samuelsson P, Caroff P and Xu H Q 2012 *Nano Lett.* **12** 228-233
[63] Kleinsasser A W, Miller R E, Mallison W H and Arnold G B 1994 *Phys. Rev. Lett.* **72** 1738–1741
[64] Ronen Y, Cohen Y, Kang J H, Haim A, Rieder M T, Heiblum M, Mahalu D and Shtrikman H 2016 *PNAS* **113** 1743-1748
[65] Abay S, Persson D, Nilsson H, Wu F, Xu H Q, Fogelström M, Shumeiko V and Delsing P 2014 *Phys. Rev. B* **89** 214508